# Spin texture on top of vortex avalanches in Nb/Al$_2$O$_3$/Co thin film heterostructures

R. F. Lopes[1], D. Carmo[2], F. Colauto[2,3], W. A. Ortiz[2], A. M. H. de Andrade[1], T. H. Johansen[4], E. Baggio-Saitovitch[5], P. Pureur[1]

[1] *Instituto de Física, Universidade Federal do Rio Grande do Sul, P.O. Box 15051, 91501-970 Porto Alegre, RS, Brazil.*

[2] *Departamento de Física, Universidade Federal de São Carlos, 13565-905 São Carlos, SP, Brazil.*

[3] *Materials Science Division, Argonne National Laboratory, Argonne, Illinois 60439, USA.*

[4] *Department of Physics, University of Oslo, P. O. Box 1048 Blindern, 0316 Oslo, Norway and Institute for Superconducting and Electronic Materials, University of Wollongong, Northfields Avenue, Wollongong, NSW 2522, Australia.*

[5] *Centro Brasileiro de Pesquisas Físicas, Rua Dr. Xavier Sigaud 150, 22290-180 Rio de Janeiro, RJ, Brazil.*

**Abstract**

We report on magneto-optical imaging, magnetization, Hall effect and magneto-resistance experiments in Nb/Al$_2$O$_3$/Co thin film heterostructures. The magnetic field is applied perpendicularly to the plane of the film and gives rise to abrupt flux penetration of dendritic form. A magnetization texture is imprinted in the Co layer in perfect coincidence with these ramifications. The spin domains that mimic the vortex dendrites are stable upon the field removal. Moreover, the imprinted spin structure remains visible up to room temperature. Complementary magnetization, Hall effect and magneto-resistance experiments were performed in a similar sample where electrical contacts were placed on the Co layer. In the region of the field - temperature diagram where flux instabilities are known to occur in Nb films, irregular jumps are observed in the magnetic hysteresis and large amplitude noise is detected in the magneto-resistance and Hall resistivity data when measured as a function of the field.

# I. INTRODUCTION

The study of non-collinear spin configurations structured at meso and microscopic scales have been a subject of great interest for fundamental and applied research. Previously, investigations were devoted to understand the microscopic mechanisms leading to non-trivial spin arrangements [1]. On the other hand, in the present time many of these structures may have potential application in innovative processes involving data manipulation and spintronics devices [2].

Domain walls [1], small magnetic domains in single or multilayered thin films with perpendicular anisotropy [3], spontaneous bubble domains formed on the surface of a bulk ferromagnetic material with planar anisotropy [4], and stripe domains in multilayered magnetic films [5] are just a few examples of spin textures structured at sub-micron scale. These structures in general result from the interplay between the exchange energy, magnetocrystalline anisotropy and dipolar interaction.

Recently, novel spin textures have been put forward. Among them are vortex- and chiral-type magnetic arrangements. Spin configurations in form of vortices structured at nanometric scale are usually associated to very small samples with circular geometries [6]. Chiral magnetic structures are extremely interesting since their stabilization is crucially dependent on the spin-orbit interaction [7]. Chiral spin textures, for instance, have been proposed to explain the anomalous Hall effect in the Kondo lattice $Pr_2Ir_2O_7$ [8] and were observed in the surface states of the topological insulator $Bi_{1-x}Sb_x$ with ARPES experiments [9].

A remarkable nanoscale chiral spin structure is the magnetic skyrmion, discovered a few years ago [10, 11]. Reference [10] reports the first observation of a skyrmion lattice using neutron scattering in the magnetic crystal MnSi that lacks the inversion symmetry. Other observations of skyrmions using different experimental techniques have been reported in bulk materials and thin films [12, 13]. Skyrmions are described as quasiparticles in an otherwise homogeneous magnetic medium that are formed by a local spin pointing opposite to the global magnetization that is surrounded by a whirling twist of spins as illustrated in figure 1 of reference [14]. These spin arrangements are stabilized by the Dzyaloshinskii-Moriya interaction, whose magnitude depends on the spin-orbit coupling [14].

Heterostructures formed by bilayers and multilayers including ferromagnetic and superconducting thin films are also potentially interesting systems to produce unusual spin textures. However, most of the works on these hybrids are dedicated to the study of the effects of the magnetic layers on the superconducting films [15]. A relatively smaller number of investigations are focused in the effects of superconductivity on the spin configuration of the magnetic layers [16, 17, 18]. One interesting possibility along this line of investigation is the generation of unconventional spin textures in the ferromagnetic layers using the properties of the superconducting films. It seems reasonable to suppose that the Abrikosov vortex lattice of type-II superconducting layers may induce nanoscopic and non-trivial spin structures in nearby ferromagnetic layers grown from materials with strong spin-orbit coupling. Owing to the hexagonal symmetry that characterizes both the Abrikosov vortex lattice [19] and the skyrmion lattice [10], one might expect that some chiral-type spin configuration forms on top of superconducting vortices, at least in the field limit where the distance between vortices is larger than the typical spin quasiparticle diameter.

In this article we report on spin texture generated by vortices in a Nb/Al$_2$O$_3$/Co thin film hybrid. The magnetic field applied perpendicularly to the superconducting layer gives rise to abrupt flux penetration in the form of dendrites. A magnetization texture is imprinted in the Co layer in perfect coincidence with these ramifications. Magneto-optical imaging (MOI) experiments were carried out to observe the flux avalanches and the imprinted spin texture. It is remarkable that the spin domains that mimic the vortex dendrites are stable upon the field removal. Moreover, the imprinted spin structure remains visible up to room temperature. Complementary magnetization, magneto-resistance and Hall effect experiments were performed in a similar sample were electrical contacts were placed on the Co film. In the region of the H-T diagram where flux instabilities are known to occur in the Nb film [20], irregular jumps are observed in the magnetic hysteresis. The magneto-resistance and the Hall resistivity show a noisy behavior when measured in fixed temperatures under slowly varying magnetic fields. The noise in the magneto-transport measurements ceases at a characteristic field which depends on the temperature. Using the magneto-resistance and Hall resistivity data we were able to define the boundary of the instability region for the vortex penetration.

## II.     MATERIAL AND METODS

Several samples of the thin film heterostructure Nb/Al$_2$O$_3$ /Co were made by UHV magnetron sputtering using the system Orion 8$^@$ manufactured by AJA International Inc. The films thicknesses for all samples are 200 nm, 22 nm and 24 nm for Nb, Al$_2$O$_3$ and Co, respectively. The thicknesses were determined by x-ray reflectomety for each film individually. Base pressure was better than 5 10$^{-8}$ Torr. Growth was done under the Ar working pressure of 2 mTorr. The surface of the metallic films is homogeneous when observed using optical microscopy. The structure of the prepared heterostructures is show in the schematic picture of Fig. 1. First, the Nb film with area 7x7 mm$^2$ was deposited on a thermally oxidized Si (100) wafer. This particular layer was grown at $T = 500\ ^0$C. Then, the Al$_2$O$_3$ layer was deposited on the Nb film covering its entire surface. Above the insulating layer, a rectangular shaped Co film with areal size 2 x 4 mm$^2$ was deposited. The Al$_2$O$_3$ spacer prevents electrical metallic contact between the superconductor and the ferromagnetic sheets. The insulator layer also avoids the proximity effect between the metallic films. On top of the Co film, a thin Al$_2$O$_3$ layer was grown in order to prevent sample degradation. Six thin and small Ag pads were deposited on the Al$_2$O$_3$ film previously to the Co film growth. In the extremities of these pads, shown in Fig. 1, gold wires were glued with silver paint. These pads provide electrical contacts to the Co film so that magneto-resistance and Hall effect experiments could be carried out in this particular layer. The Ag contact pads and the rectangular Co were shaped using a hollowed mask placed on the samples during the deposition process. In some samples, additional electrical contacts could be placed in order to test the superconducting layer only. The Nb films of our heterostructures have a critical temperature $T_c$ = 8.9 K,

The magnetization measurements were performed in a Quantum Design MPMS-5S@. The field was applied perpendicularly to the plane of the film. The images were carried out in a magneto-optical imaging (MOI) setup, using a Faraday rotating Bi-substituted ferrite-garnet film with in-plane magnetization as a sensor [21, 22]. Mounting the sensor plate directly on the film surface, the flux density distribution over the film area was detected as an image in a polarized light microscope with crossed polarizers. Electrical resistivity, magneto-resistance and Hall effect measurements were carried out in a Quantum Design DynaCool@ Physical Properties Measuring System. In the magneto-transport experiments, the magnetic field was oriented perpendicular to the film surface.

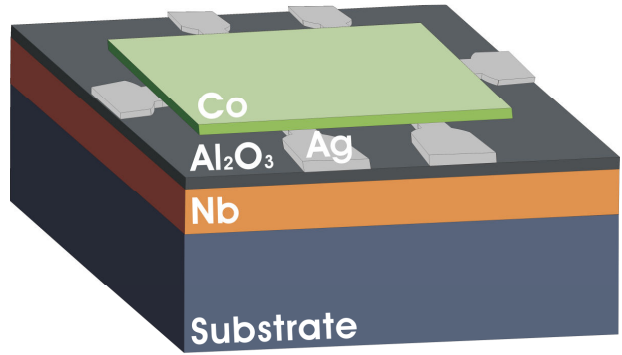

FIG. 1. Studied Nb/Al$_2$O$_3$/Co heterostructure (schematic). The Si$_{100}$ substrate has 400 μm thickness. The material layers and thicknesses are in the following sequence: Nb, 200 nm; Al$_2$O$_3$, 22 nm; Ag, 20 nm; Co, 24 nm; and not shown Al$_2$O$_3$, 10 nm.

### III. RESULTS AND DISCUSSION

III.a *Magnetization and magneto-optical measurements*

It is well known that Nb films exhibit flux avalanches when submitted to magnetic fields within a certain range and for temperatures below 5 K approximately [20]. We have confirmed this feature in our sample, as can be observed in Fig. 2 that shows hysteresis cycles in fields up to $B$ = ± 20 mT at four different temperatures. Before each isothermal cycle the sample was warmed up above $T_c$. At this stage, the Co layer was demagnetized by successively applying fields with alternating orientations and decreasing amplitude. The isotherms at 6 K and 8 K are smooth, as expected for a superconductor in the critical state. Consistently, the area of the hysteresis loop at 6 K is larger than that registered at 8 K. On the other hand, the isotherms at 2 K and 4 K differ from the ordinary behavior, having an areal size smaller than that in 6 K. In addition, they look strongly noisy, both in the upward and downward branches, which is a well-known signature of flux entry in the form of avalanches [20, 23].

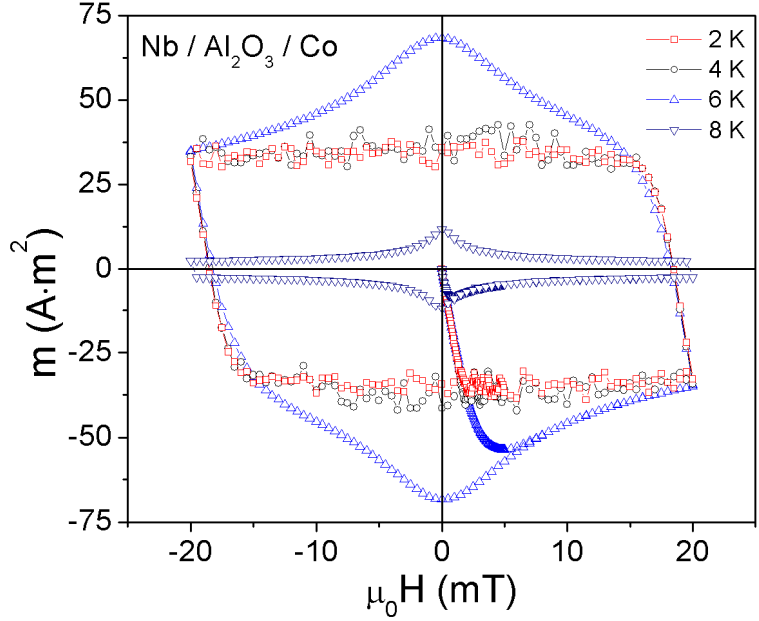

FIG. 2. Magnetic moment versus applied field for the heterostructure Nb/Al$_2$O$_3$/Co. The noise in the hysteresis loop at $T = 2$ K and $T = 4$ K is typical of the occurrence of flux avalanches.

The avalanches develop inside the sample with a dendritic morphology, as observed in the magneto-optical image in Fig 3 (a) and (b). Since the sample is larger than the microscope field of view, the magneto-optical images were taken in a fraction of the sample, which encompasses the edge between the bare Nb and the Nb/Al$_2$O$_3$/Co heterostructure. Vertical lines were added to the figure in order to identify the boundary separating the uncovered Nb layer and the heterostructure. The images (a) and (b) were carried out after a zero field cooling (ZFC) from a temperature above $T_c$ to 2.5 K and the image (c) shows the remnant out-of-plane magnetization of the Co film. In the image (a), taken at 2.5 mT, one sees two dendritic structures that start at the edge of the sample and reach the area including the magnetic layer. There is no significant deflection or suppression of branches as the avalanche invades the area covered by the Co layer. However, those branches that are about to end at the boundary are stopped. When the field is increased to 6.0 mT, as shown in image (b), several dendrites penetrate inside the heterostucture. When the magnetic field is removed and the sample is warmed to a temperature above $T_c$ the Co layer stays magnetized as is shown in the image (c). Through a meticulous comparison between the images (b) and (c), one sees that the same branches developed at 2.5 K are imprinted in the magnetic layer.

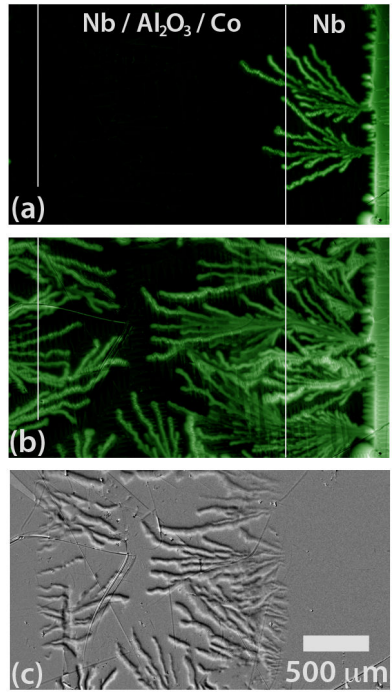

FIG. 3. Magneto-optical imaging (a) 2.5 K and 2.5 mT; (b) 2.5 K and 6 mT; (c) Remnant state of the Co layer at $T = 10$ K.

The magnetic texture produced by flux avalanche remains stable in the Co layer in temperatures much above $T_c$ and is erased only by applying a magnetic field of 1 T magnitude or higher. As an example, Fig. 4 shows a micrograph of imprinted vortex obtained in room temperature. Although some branches merged into a bulky one and others look blurry at this temperature, their observation is an indication of the robustness of the imprinted magnetic structure.

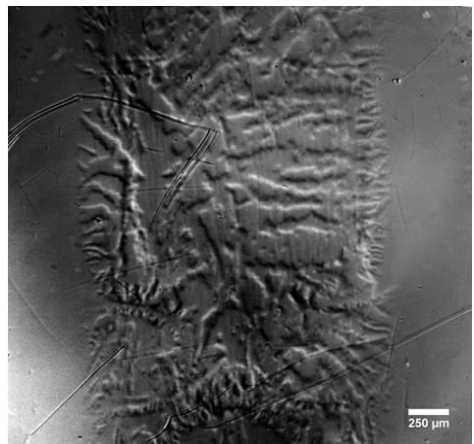

FIG. 4. Magneto-optical imaging of the Co layer showing imprinted vortex avalanches in $T = 300$ K . Bar corresponds to 250 μm.

The vortex induced spin texture also affects the field penetration in temperatures where the field penetrates smoothly, as can be observed in Fig. 5, which shows a MOI experiment carried out at $T = 8$ K. It is quite visible the difference between the bare Nb and the Nb/Al$_2$O$_3$/Co regions. In the uncovered Nb film the flux distribution is homogeneous, while in the Nb/Al$_2$O$_3$/Co area the flux is guided by the previous dendrite imprints. The image (a) was taken at 0.3 mT. It shows the beginning of flux penetration into the Co area. Increasing the field, as in image (b) taken at 0.6 mT, the flux occupies the locations magnetized by the avalanches. Clearly, the flux follows the branches and bifurcations left by dendrites. The image (c) was taken at 1 mT. It shows the surface appearance when the flux is almost fully penetrated into the sample. The flux inhomogeneity takes all over the Nb/Al$_2$O$_3$/Co portion that was previously submitted to avalanche occurrences at a lower temperature.

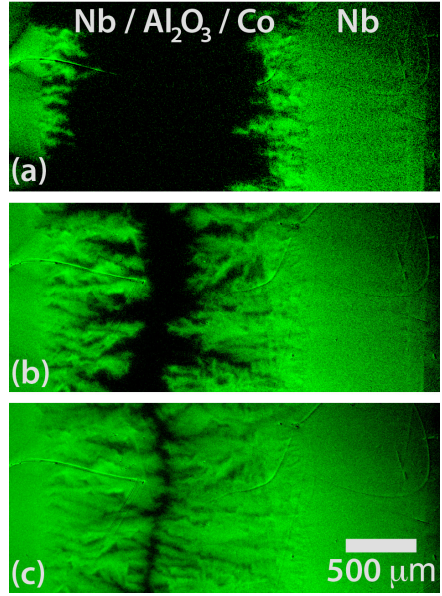

FIG. 5. Magneto-optical imaging at $T$ = 8K (a) 0.3 mT; (b) 0.6 mT; (c) 1 mT.

III.b  *Hall effect and magneto-resistance experiments*

In this section, magneto-transport measurements in the Nb/Al$_2$O$_3$/Co hybrid are presented and discussed. The results shown are those obtained when the electrical contacts were placed on the ferromagnetic film. We notice that the longitudinal resistivity (results not show) exhibit a metallic behavior, as usually observed in cobalt thin films [25]. The measured residual resistivity is $\rho_0 \approx 15\mu\Omega$cm, which is a typical value for a Co film with 20 nm thick [25, 26]. Moreover, the overall behavior of our magneto-transport data in Co at temperatures above the superconducting $T_c$ of the nearby Nb layer is in agreement with previously reported results [25-27].

Figure 6 shows representative Hall resistivity results in low fields at some fixed temperatures below and above $T_c$. In $T$ = 2 K and $T$ = 4 K a rather noisy behavior is observed in $\rho_{xy}$ in the low applied field limit. The noise amplitude is much larger than the experimental resolution of the PPMS$^@$. The erratic fluctuations in $\rho_{xy}$ are sharply suppressed above a temperature-dependent threshold field $\mu_0H_t(T)$, signaled by arrows in Fig. 6. The values obtained for $\mu_0H_t$ in different isotherms are plotted in the diagram of Fig. 8. These field values define an upper boundary for observing instabilities in the magneto-transport properties. As shown in Fig. 6(c), the noise is not observed in measurements carried out above $T_c$. We attribute the observed instabilities in $\rho_{xy}(H)$ to successive voltage spikes induced in orientations parallel to the Co layer by the abrupt penetration of the vortex dendrites. Indeed, since the average time for avalanche penetration ranges around 0.1 to 0.4 µs [28, 29], the penetration of a single bundle of 100 vortices may produce an easily detectable spike of 1 µV magnitude.

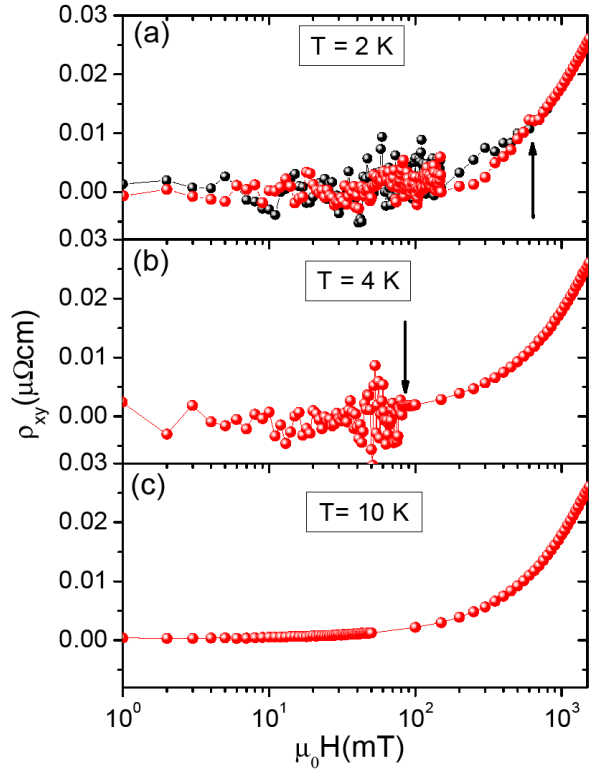

FIG. 6. Low field Hall resistivity as a function of the applied field of the Co layer in the Nb/Al$_2$O$_3$/Co hybrid measured in temperatures (a) 2 K, (b) 4 K e (c) 10 K. Red balls represent results recorded while the field is increased and black symbols correspond to data obtained in decreasing fields. Arrows denote the maximum field for observing instabilities in the $\rho_{xy}$ data.

Magneto-resistance (MR) measurements versus the applied field in the same fixed temperatures studied in the $\rho_{xy}$ experiments are shown in Fig. 7. The MR shows a noisy behavior in the same field and temperature intervals as for the Hall resistivity data. In addition, the usually observed hysteretic behavior in the MR of ferromagnetic metals is rounded off in the temperature range where instabilities were identified in the Co layer of our hybrid. As seen in Fig. 7(c), the normal hysteretic behavior of the MR is restored above $T_c$ [25]. The peaks indicate the position of the coercive field $\mu_0 H_c \cong 0.05$ T. The non-observation of hysteresis in the MR results at $T = 2$ K (see Fig. 7(a)), is probably due to a smoothing resulting from rapid inductive fluctuations due to vortex avalanches. In $T = 4$ K (Fig. 7(b)), the peaks at $\pm H_c$ are discernible though smudged by large amplitude noise.

As for the Hall resistivity, the anomalous fluctuations in $\rho_{xx}$ cease suddenly at the threshold field $\mu_0H_t(T)$. The values for this characteristic field obtained from the MR measurements are also plotted in Fig. 8.

It is worthwhile to note that a simple device as our trilayer heterostructure allows the detection of vortex avalanches using a conventional electrical transport measurement technique. Clearly, in the present stage, the signal detection method employed in our measurements does not allow detailed studies of the noise spectrum as done in the more complex experiments reported in references [28] and [29]. However, the use of a normal layer inductively coupled to an electrically isolated superconducting film might be an interesting alternative to study the rapid dynamics of vortex penetration.

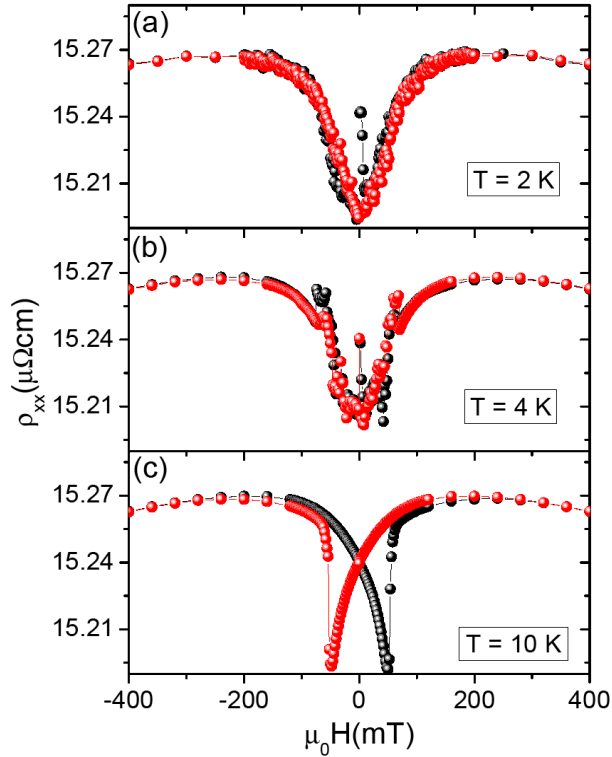

FIG. 7. Low field magneto-resistance of the Co layer in the Nb/Al$_2$O$_3$/Co hybrid measured in temperatures (a) 2 K, (b) 4 K and (c) 10 K. Black balls represent measurements taken as the field increases from $\mu_0H = -0.4$ T up to $\mu_0H = +0.4$ T. The data plotted as red balls were obtained while the field was varied in the opposite way.

In the diagram of Fig. 8 the threshold induction $\mu_0H_t$ is plotted as a function of the temperature in a logarithmic scale for better visualization of the results in temperatures close to $T_c$. We propose that this characteristic field defines the upper boundary for the region where vortex instabilities occur in the Nb layer of our heterostructure. Indeed, for field values above 10 mT and temperatures below 5 K, the diagram in Fig. 8 reproduces quite nicely similar results reported for Nb films with thicknesses varying from 20 to 500 nm [20]. In the low $H$ – high $T$ region, the diagram in Fig. 8 does not show the reentrance of the instability region in the single layer Nb [20]. Figure 8 suggests instead that $\mu_0 H_t(T)$ approaches zero near $T_c$. We note that similar enlargements of the *H-T* region characterized by irregular jumps in magnetization measurements were observed in Nb/permalloy bilayers [30]. Authors in Ref. [30] suggest that the pinning of vortices in channels due to the underlying magnetic domains combined with the intrinsic pinning of the Nb film could explain the enlargement of the *H-T* domain where vortex instabilities are observed.

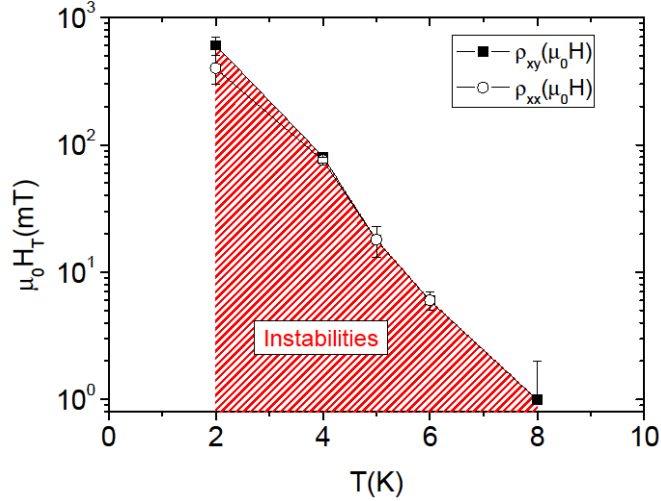

FIG. 8 – Characteristic magnetic induction $\mu_0 H_t(T)$ plotted as a function of the temperature for the Nb/Al$_2$O$_3$/Co hybrid. The *H* versus *T* line drawn through the experimental points denotes the upper limit for observing noisy behavior in the Hall resistivity and magneto-resistance measurements. This line corresponds to the instability boundary for vortex penetration into the Nb layer.

## IV.    FINAL CONSIDERATIONS AND CONCLUSION

Our magneto-optical experiments in Nb/Al$_2$O$_3$/Co heterostructures clearly show that dendrites produced by vortex avalanches in the Nb layers generate a template consisting of magnetic textures in the adjacent Co layer. The dynamics of the avalanches induces anomalous noise in magneto-transport properties of the Co film. Magnetization measurements in these tri-layer heterostructures show irregular jumps consistently both with MOI experiments and magneto-transport results.

Due to the similarities between the vortex lattice in superconductors and the skyrmion lattice in magnetic materials, we are lead to consider that the macro and mesoscopic magnetization texture observed in the Co layer of our Nb/Al$_2$O$_3$/Co hybrid might be microscopically formed by skyrmions or some other non-conventional spin chiral or vortex-type structures. This hypothesis, however, should be checked by performing other experiments able to discern magnetization inhomogeneities at the nanoscopic scale. At this stage, we can not exclude the possibility that the spin texture clearly revealed in Fig. 3(c) is related to perpendicular domain polarization following the geometry dictated by the flux penetration into the Nb layer. However, it was shown with magnetic force microscopy that in Co films having 20 nm thickness, which is close to that of the Co layer in our hybrid, only domains with in-plane magnetization are stable [25]. Thus, it might be difficult to conceive that small domains with perpendicular magnetization would subsist in the absence of vortices and applied fields in the Co layer of the studied heterostructure. Moreover, a rough estimation shows that the magnetic energy density transferred by vortices to the ferromagnetic film is not enough to overcome the in-plane shape anisotropy energy for a thin Co film, which is of the order $K = 1.3 \; 10^6$ J/m$^3$, as reported in Ref [31].

As a final remark, we mention that the imprinting of vortex avalanches in a ferromagnetic thin film might represent a step forward to the development of an alternative technique for achieving vortex decoration in a medium useful for data recording and processing.

**Note added**.

After the completion of this manuscript we become aware of a recent work by J. Brisbois et al. [32] where observations of vortex avalanches imprints in a permalloy film adjacent to a Nb layer are reported.

**Acknowledgements.**

We acknowledge Dr. Eduardo M. Bittar from the Centro Brasileiro de Pesquisas Físicas (CBPF) for invaluable help during the realization of the magneto-transport experiments. One of the authors (RFL) acknowledges the Brazilian agency "Conselho Nacional de Pesquisas Científicas e Tecnológicas" (CNPq) for financial support.